\documentclass[fleqn,11pt,twoside]{article}

\usepackage{amsthm,amsthm,amssymb, color, xcolor,epsfig, graphics, subfigure}

\usepackage{amsmath, graphicx, latexsym, lscape }

\makeatletter
\newcommand{\copyrightnote}[2]{{\renewcommand{\thefootnote}{}
 \footnotetext{\small\it
\begin{flushleft}
 \copyright \ #1   #2  
\end{flushleft}}}}

\newcommand{\Name}[1]{\begin{flushleft}
                       \LARGE \bf #1
                       \end{flushleft}\vspace{-3mm}}

\newcommand{\Author}[1]{\begin{flushleft}
                       \it #1 \end{flushleft}}

\newcommand{\Address}[1]{\begin{flushleft}
                       \it #1 \end{flushleft}}

\newcommand{\Date}[1]{\begin{flushleft}
                      \small  \it #1 \end{flushleft}}

\newcommand{\evenhead}{Author \ name}
\newcommand{\oddhead}{Article \ name}

\renewcommand{\@evenhead}{
\hspace*{-3pt}\raisebox{-15pt}[\headheight][0pt]{\vbox{\hbox to \textwidth
{\thepage \hfil \evenhead}\vskip4pt \hrule}}}
\renewcommand{\@oddhead}{
\hspace*{-3pt}\raisebox{-15pt}[\headheight][0pt]{\vbox{\hbox to \textwidth
{\oddhead \hfil \thepage}\vskip4pt\hrule}}}
\renewcommand{\@evenfoot}{}
\renewcommand{\@oddfoot}{}

\long\def\@makecaption#1#2{%
  \vskip\abovecaptionskip
  \sbox\@tempboxa{\small \textbf{#1.}\ \ #2}%
  \ifdim \wd\@tempboxa >\hsize
    {\small \textbf{#1.}\ \ #2}\par
  \else
    \global \@minipagefalse
    \hb@xt@\hsize{\hfil\box\@tempboxa\hfil}%
  \fi
  \vskip\belowcaptionskip}

\newcommand{\JNMPnumberwithin}[3][\arabic]{%
  \@ifundefined{c@#2}{\@nocounterr{#2}}{%
    \@ifundefined{c@#3}{\@nocnterr{#3}}{%
      \@addtoreset{#2}{#3}%
      \@xp\xdef\csname the#2\endcsname{%
        \@xp\@nx\csname the#3\endcsname .\@nx#1{#2}}}}%
}

\newcommand{\resetfootnoterule} {
  \renewcommand\footnoterule{%
  \kern-3\p@
  \hrule\@width.4\columnwidth
  \kern2.6\p@}
}

\renewcommand{\footnoterule}{}

\makeatother

\theoremstyle{definition}

\begin{document}

\renewcommand{\evenhead}{ {\LARGE\textcolor{blue!10!black!40!green}{{\sf \ \ \
]ocnmp[}}}\strut\hfill Aratyn, Gomes, Lobo, Zimerman}
\renewcommand{\oddhead}{ {\LARGE\textcolor{blue!10!black!40!green}{{\sf ]ocnmp[}}}
\ \ \ \ \ Painlev\'e  $D_4^{(1)}$ system in a self-similarity limit}

\thispagestyle{empty}
\newcommand{\FistPageHead}[3]{
\begin{flushleft}
\raisebox{8mm}[0pt][0pt]
{\footnotesize \sf
\parbox{150mm}{{Open Communications in Nonlinear Mathematical Physics}\ \  \ {\LARGE\textcolor{blue!10!black!40!green}{]ocnmp[}}
\ \ Vol.3 (2023) pp
#2\hfill {\sc #3}}}\vspace{-13mm}
\end{flushleft}}

\FistPageHead{1}{\pageref{firstpage}--\pageref{lastpage}}{ \ \ Article}

\strut\hfill

\strut\hfill

\copyrightnote{The author(s). Distributed under a Creative Commons Attribution 4.0 International License}

\Name{Derivation of Painlev\'e  type system 
with $D_4^{(1)}$ affine Weyl group symmetry in a self-similarity limit
}

\Author{H. Aratyn$^{\,1}$, J.F. Gomes$^{\,2}$ , G.V. Lobo$^{\,2}$
and A.H. Zimerman$^{\,2}$}


\Address{$^{1}$ Department of Physics, 
University of Illinois at Chicago, 
845 W. Taylor St.
Chicago, Illinois 60607-7059, USA\\[2mm]
$^{2}$ Instituto de F\'{\i}sica Te\'{o}rica-UNESP,
Rua Dr Bento Teobaldo Ferraz 271, Bloco II,
01140-070 S\~{a}o Paulo, Brazil }

\Date{Received August 9, 2023; Accepted September 1, 2023}

\setcounter{equation}{0}

\begin{abstract}
\noindent 
We show how the zero-curvature equations based on a loop algebra 
of $D_4$ 
with a principal gradation reduce via self-similarity limit
to a polynomial Hamiltonian 
system of coupled Painlev\'e III models with four canonical variables
and $D_4^{(1)}$ affine Weyl group symmetry.
\end{abstract}

\label{firstpage}


\section{Zero curvature derivation of the $t_3$  flow of the
$D_4^{(1)}$ hierarchy}
Our starting point will be an integrable hierarchy with commuting flows 
defined via the zero-curvature formalism based on a loop algebra of 
${\cal G}=D_4$ endowed
with a principal gradation. 
We will apply a conventional self-similarity limit on the $t_3$ flow
of the hierarchy and show how through several explicit 
changes of variables 
the reduced model can be cast into the polynomial 
Hamiltonian system of four canonical variables invariant under 
$D_4^{(1)}$ affine Weyl group symmetry. 

An unusual aspect of the  $t_3$ flow of the $D_4$ hierarchy is that it is 
parametrized by two independent variables, $\varepsilon_1,
\varepsilon_2$, reflecting the
fact that a kernel of $E^{(1)}$, a central object of zero-curvature
equations, turns out to have  a  two-dimensional kernel on the level
of grade three. The presence of these parameters enriches the symmetry
structure of the two-dimensional
hierarchy of zero-curvature equations based on $D_4^{(1)}$ affine Weyl
algebra and survives the self-similarity limit as shown
in equations \eqref{viscuvx20ss}.
Considering these equations with dependence on only one of these 
parameters or their linear combination we obtain, up to few normalization adjustments,
and after several changes of variables, the model of 
reference \cite{sasanoD41}, where it was proposed
as a pair of coupled Painlev\'e III equations that form
a Hamiltonian system invariant under 
$D_4^{(1)}$ affine Weyl group symmetry.
To the best of our knowledge, an explicit 
derivation of this particular 
model as a reduction of two-dimensional
hierarchy of zero-curvature equations based on $D_4^{(1)}$ affine Weyl
algebra, has not been done previously.

To make the presentation self-contained we 
provide all the necessary algebraic background information in
Appendix \ref{section:algebraic} and main expressions of the
zero-curvature calculation in Appendix \ref{section:curvaturanula}.

For other derivations of integrable hierarchy based on $D_4^{(1)}$ affine
algebra the reader  can consult \cite{gerdjikov} and the
references therein.
There also exist in the literature other approaches to applying 
similarity reduction to integrable hierarchy of type $D_4^{(1)}$ 
\cite{FS} but the focus there was on deriving the sixth Painlev\'e equation.

As a starting point we consider the zero-curvature equation for the third flow
in the setting of the affine algebra $D_4^{(1)}$:
\begin{equation}
 [ \partial_x +E^{(1)} +A_0\,,\, \partial_{t_3} + D^{(0)}+
 D^{(1)}+D^{(2)}+D^{(3)} ]=0
\label{zerocurva}
\end{equation}
with $D^{(i)} \in {\cal G}_i$ and $A_0=\sum_{i=1}^4 \phi_i H_i $ with
$E^{(1)}, {\cal G}_i,H_i$ defined in Appendix \ref{section:algebraic}.

The highest grade-four  component of equation \eqref{zerocurva}, 
$\lbrack E^{(1)}, D^{(3)}
\rbrack=0$, is solved by
\begin{equation}
D^{(3)}=  \varepsilon_1 V_1+\varepsilon_2V_2\, ,
\label{D3}
\end{equation}
where $V_1, V_2$ are two matrices defined in \eqref{e1k3} that 
span a basis for the two-dimensional kernel of $E^{(1)}$ in ${\cal
G}^{(3)}$.
The standard zero-curvature technique allows recursive derivation
of the lower grade matrices $D^{(i)}, i=0,1,2$
from appropriate grade projections of equation 
\eqref{zerocurva}.
Grade $2$ element given by:
\begin{equation}
 D^{(2)}= M_1 E^{(0)}_{e_1-e_3} +M_2  E^{(0)}_{e_2+e_4} +M_3
E^{(0)}_{e_2-e_4}+ M_4  E^{(1)}_{-e_1-e_3} \, ,
 \label{DtwoM}
\end{equation}
is solved for from the grade $3$ equation
\begin{equation}
 \lbrack  E^{(1)}, D^{(2)} \rbrack + \lbrack A_0, 
D^{(3)}\rbrack= 0\, ,
\label{grade3eq}
 \end{equation}
from which one obtains explicitly coefficients $M_i, i=1,2,3,4$ 
of $ D^{(2)}$ given in equation \eqref{Mphis}.

The grade $2$ component of the zero curvature equations \eqref{zerocurva}
is
\begin{equation}
\lbrack  E^{(1)}, D^{(1)} \rbrack + \lbrack A_0, 
D^{(2)}\rbrack  +\partial_x D^{(2)}= 0 \, , 
\label{grade2curva}
 \end{equation}
where
\begin{equation}
D^{(1)}=  d_1 E_1+d_2 E_2+d_3 E_3+d_4 E_4+d_5 E_5 \, , 
\label{Done1}
\end{equation}
where we employed the basis elements $E_i, i=1,{\ldots}, 5$ given in
expressions \eqref{g1basis}. Equation \eqref{grade2curva} 
yields  explicit expressions for $d_i,i=2,{\ldots} ,5$ given in 
equation \eqref{dphis}.

The grade one component of \eqref{zerocurva} reads as
\begin{equation}
\partial_x D^{(1)} +\lbrack E^{(1)} , D^{(0)} \rbrack
+\lbrack A_0 , D^{(1)} \rbrack=0\, ,
\label{gradeone}
\end{equation}
with $D^{(0)} = \sum_i v_i H_i$.
The advantage of using the basis \eqref{g1basis} is that
\begin{equation}
\begin{split}
\lbrack E^{(1)} , D^{(0)}  \rbrack&= (v_2-v_1)E^{(0)}_{e_1-e_2}+
(v_3-v_2) E^{(0)}_{e_2-e_3}+ (v_4-v_3) 
E^{(0)}_{e_3-e_4}\\&- (v_3+v_4)E^{(0)}_{e_3+e_4}+(v_1+v_2)
E^{(1)}_{-e_1-e_2}\\
&= v_1 E_2 + v_2 E_3 + v_3 E_4 + v_4 E_5\, ,
\label{E1D0}
\end{split}
\end{equation}
where $E_i, i=1,{\ldots} E_5$ are the basis elements given in
expressions \eqref{g1basis}.

Solving the grade one equation \eqref{gradeone} in direction of $E_1$ yields
\begin{equation}
\partial_x d_1 = -\frac13 \phi_2d_4-\frac13 \phi_3d_3+\frac13 \phi_1d_2+\frac13 \phi_4d_5+\frac23 \phi_3d_4
 +\frac23 \phi_2d_3,
\label{bd1}
\end{equation}
plugging expressions \eqref{dphis} and taking out the total
derivative gives
\begin{equation}
d_1 =  \frac{\varepsilon_1+\varepsilon_2}{12} \left( 
-\phi_1^2-\phi_4^2 +  \phi_2^2+\phi_3^2 +2 \phi_2\phi_3\right)
+  \frac{1}{6} (\varepsilon_1-\varepsilon_2)  \phi_1 \phi_4 \, .
\label{d1}
\end{equation}
Solving the grade one equation \eqref{gradeone} in directions of $E_i,
i=2,3,4,5$ yields
\begin{equation} 
v_i= - \partial_x d_{i+1}-C_{i+1} , \qquad i=1,2,3,4 \, , 
\label{vidici}
\end{equation}
with $C_i, i=2,3,3,5$ given in \eqref{newCia} and with $d_1$ in given in
expression  \eqref{d1} while expressions $d_i, i=2,..,5$ are 
given in \eqref{dphis}.

Inserting these values of $d_i$ and $C_i$ into \eqref{vidici} we obtain
$v_i$ given in expression \eqref{vis}.

The grade zero component of \eqref{zerocurva} is 
\begin{equation}
\partial_x D^{(0)} - \partial_{t_3} A_0 +\lbrack A_0 , D^{(0)}
\rbrack=0\, .
\label{gradezero}
\end{equation}
Since $\lbrack A_0 , D^{(0)} \rbrack=0$ 
the equation \eqref{gradezero} reduces to
\[
\partial_{t_3} A_0 =  \sum_i \partial_{t_3} \phi_i H_i= \partial_x D^{(0)} 
= \sum_i \partial_x v_i H_i \, ,
\]
that in components gives the $t_3$ flows written as
\begin{equation}
\partial_{t_3} \phi_i=
\partial_x v_i, \; i=1,2,3,4\, .
\label{t3flows}
\end{equation}
When on the right hand side we insert values of $v_i$ from equation 
\eqref{vis}  we find the $t_3$-flow explicitly written in equation
\eqref{visb} with their symmetries listed below in equations 
\eqref{symmetryso8}.

With definitions
\begin{equation} 
u = \phi_1 + \phi_4,\; v = \phi_1 - \phi_4, 
\;
f = \phi_2 +\phi_3, \;
g= \phi_2 - \phi_3 \,. \;
\label{Phipmdefs}
\end{equation}
Equations \eqref{visb} can be conveniently rewritten in this notation as
\begin{equation}
\begin{split}
\partial_{t_3} u &= \varepsilon_1 \partial_x(\frac14 u v^2 - \frac14 u f^2 +
\frac12 v \partial_x f - \frac12 f \partial_x v)
\\&+\varepsilon_2 \partial_x(\frac14 u v^2 - \frac14 u g^2 +v \partial_x f
+\frac12 u \partial_x g + \frac12 f \partial_x v+\partial_x^2 u)\, ,
\\
\partial_{t_3} v &= \varepsilon_1 \partial_x(\frac14 v u^2 - \frac14 v g^2 
+ u \partial_x f +\frac12 v \partial_x g + \frac12 f \partial_x
u+\partial_x^2 v)\\&+\varepsilon_2 \partial_x(\frac14 v u^2 - \frac14 v f^2 
+ \frac12 u \partial_x f - \frac12 f \partial_x
u)\, ,\\
\partial_{t_3} f &= \varepsilon_1 \partial_x(\frac14 f g^2 - \frac14 f u^2 
- u \partial_x v -\frac12 v \partial_x u - \frac12 f \partial_x
g-\partial_x^2 f)\\
&+\varepsilon_2 \partial_x(-\frac14 f v^2 + \frac14 f g^2 
- v \partial_x u -\frac12 u \partial_x v - \frac12 f \partial_x
g-\partial_x^2 f)\, ,\\
\partial_{t_3} g &= \varepsilon_1 \partial_x(\frac14 g f^2 - \frac14 g v^2 +
\frac12 f \partial_x f - \frac12 v \partial_x v) \\&+\varepsilon_2 \partial_x(\frac14 g f^2 - \frac14 g u^2 
+\frac12 f \partial_x f - \frac12 u \partial_x u)\, .
\label{viscuv}\end{split}
\end{equation}
These equations are invariant under:
\begin{equation}\begin{split}
F_1 &: u \leftrightarrow -v,\; \varepsilon_1 \to \varepsilon_2, \; g
\to g, f \to f\,, \\
F_4 &: u \leftrightarrow v, \; \varepsilon_1 \to \varepsilon_2, \; g
\to g, f \to f\,.
\label{symmetryso8}
\end{split}
\end{equation}
In addition for $\varepsilon_2=0$ these equations are invariant under:
\begin{equation}
F_2 : f \leftrightarrow v,\; \varepsilon_1 \to -\varepsilon_1, \; g \to g, u \to u
\label{f2def}
\end{equation}
while for $\varepsilon_1=0$ they are invariant under:
\[
F_3 : f \leftrightarrow u,\; \varepsilon_2 \to -\varepsilon_2, \; g \to g, v 
\to v\, .
\]
Obviously, for one of the parameters $\varepsilon_1$ or
$\varepsilon_2$ being zero the remaining parameter can be absorbed 
by redefining $t_3$.

The above operations satisfy $F_2^2=I=F_3^2$ and
\[ F_1F_4=F_4F_1=TG=GT: u \rightarrow -u,\; v\rightarrow -v, \; g \to g, f \to f\, , 
\]
where $T$ and $G$ are automorphisms :
\[
T : u \leftrightarrow u,\; v\to -v, \; g \to g, f \to -f\, , 
\]
\[
G : u \leftrightarrow -u ,\; v \to v, \; g \to g, f \to -f\, , 
\]
with $ F_4 T  =TF_1$.
All the above automorphisms of equations \eqref{viscuv} 
are ``mirror automorphisms'', meaning that they square to one.

\section{Self-similarity reduction for the $t_3$ flow  }

We will look at the self-similar reduction of equation \eqref{viscuv} with
\begin{equation}
\phi ( x,t )= t^{-\frac13} \varphi ( z ), \; 
z= \frac{x}{t^{\frac13}} = x t^{-1/3}\, ,
\label{selfsim}
\end{equation}
and correspondingly 
\[
\frac{d}{d x}=  \frac{d}{d z} \frac{d z}{d x} =   t^{-1/3}
\frac{d}{d z} ,\quad \frac{d}{d t}=  \frac{d}{d z} \frac{d z}{d t} = - \frac13 t^{-1}   
 \frac{d}{d z} z\, ,
\]
such that the KdV type of expression :
\[
\frac{d}{d t} \phi ( x,t ) + \beta_1   \frac{d}{d x} (\phi ( x,t ) 
\frac{d}{d x} \phi ( x,t )) +\beta_2   \frac{d}{d x} \phi^3 ( x,t )  
+ \beta_3 \frac{d^3}{d x^3}  \phi ( x,t ) =0 \, ,
\]
is transformed to an equation fully expressible in terms of functions of $z$:
\[
 \frac{d}{d z} (z \varphi ( z )) -3 
 \beta_1  \frac{d}{d z} (\varphi( z ) \frac{d}{d z} \varphi ( z ))
  -3  \beta_2  \frac{d}{d z} \varphi^3( z ) 
 -3 \beta_3 \frac{d^3}{d z^3}   \varphi ( z ) =0\, .
 \]
Following these rules we are now able to take self-similarity limit 
of equations \eqref{viscuv}  to obtain: 
\begin{equation}
\begin{split}
-\frac{z}{3} u+C_1 &= \varepsilon_1 (\frac14 u v^2 - \frac14 u f^2 +
\frac12 v \partial_z f - \frac12 f \partial_z v)\\
&+ \varepsilon_2 (\frac14 u v^2 - \frac14 u g^2 +v \partial_z f
+\frac12 u \partial_z g + \frac12 f \partial_z v+\partial_z^2 u)
\\
-\frac{z}{3} v+K_2 &= \varepsilon_1 (\frac14 v u^2 - \frac14 v g^2 
+ u \partial_z f +\frac12 v \partial_z g + \frac12 f \partial_z
u+\partial_z^2 v)\\
&+ \varepsilon_2 (\frac14 v u^2 - \frac14 v f^2 
+ \frac12 u \partial_z f - \frac12 f \partial_z
u)\\
-\frac{z}{3}  f+K_1 &= \varepsilon_1 (\frac14 f g^2 - \frac14 f u^2 
- u \partial_z v -\frac12 v \partial_z u - \frac12 f \partial_z
g-\partial_z^2 f)\\&+\varepsilon_2 (-\frac14 f v^2 + \frac14 f g^2 
- v \partial_z u -\frac12 u \partial_z v - \frac12 f \partial_z
g-\partial_z^2 f)\\
-\frac{z}{3} g +C_2 &= \varepsilon_1 (\frac14 g f^2 - \frac14 g v^2 +
\frac12 f \partial_z f - \frac12 v \partial_z v)\\
&+\varepsilon_2 (\frac14 g f^2 - \frac14 g u^2 
+\frac12 f \partial_z f - \frac12 u \partial_z u)\, ,
\label{viscuvx20ss}\end{split}
\end{equation}
where $C_i, K_i, i=1,2$ are integration constants.

Here we comment that it is enough to chose any direction in
$\varepsilon_1-\varepsilon_2$ plane 
because of a presence of previously
noticed automorphisms that establish an equivalence (by substitution) 
between any  of the one-parameter  $\varepsilon$ models in
a self-similarity limit.

For example, we notice a symmetry between $\varepsilon_2=0$ limit of
equation \eqref{viscuvx20ss}
and $\varepsilon_1=0$  limit of
equation \eqref{viscuvx20ss} via
\begin{equation}
\begin{split}
\varepsilon_2 &\longleftrightarrow \; \varepsilon_1, \quad 
(v,K_2) \longleftrightarrow \; (u,C_1), \quad
(u,C_1) \longleftrightarrow \; (v, K_2)\\
(f, K_1) &\longleftrightarrow \;(f,K_1) , \quad 
(g,C_2) \longleftrightarrow \;(g,C_2)
\end{split}
\label{x1zerox2}
\end{equation}
These substitutions follow from to symmetries $F_1,F_4$ from 
equation \eqref{symmetryso8}. Note that equations \eqref{viscuvx20ss}  with 
arbitrary $\varepsilon_1,\varepsilon_2$ remain invariant under 
transformations \eqref{x1zerox2} that interchange $u$ and $v$. 

We further point out that symmetry extends to any direction in the 
$\varepsilon_1-\varepsilon_2$ plane. For example
we can transform the system of equations 
\eqref{viscuvx20ss} with $\varepsilon_1=0$ into the system of equations  
 \eqref{viscuvx20ss} with $\varepsilon_1+\varepsilon_2=0$
with only the parameter $\varepsilon$ such that
$\varepsilon= \varepsilon_1-\varepsilon_2$ as follows
\begin{equation}
\begin{split}
\varepsilon_1=0    &\longleftrightarrow \;  \; 
\varepsilon_1+\varepsilon_2=0, \quad \varepsilon_2 
\longleftrightarrow \; \frac{\varepsilon}{2}, \\
(v,K_2) &\longleftrightarrow \; (f,K_2), \quad (f, K_1) 
\longleftrightarrow \;(u,K_1)\\
(u,C_1) &\longleftrightarrow \; (v, C_1), \quad
(g,C_2) \longleftrightarrow \;(g,C_2)\, .
\end{split}
\label{x1zerox1mx2}
\end{equation}
Thus for simplicity we will from now on only consider the self-similarity limit 
for the case of $\varepsilon_2=0$ rewritten as:
\begin{equation}
\begin{split}
-\frac{z}{3}  v_{+}+K_{+} &= \varepsilon (\frac14 v_{-} (u^2 - g^2 )
- u \partial_z v_{-} +\frac12 v_{-} \partial_z g - \frac12 v_{-} \partial_z
u+\partial_z^2 v_{-})\\
-\frac{z}{3}  v_{-}+K_{-} &= \varepsilon (\frac14 v_{+} (u^2 - g^2 )
+ u \partial_z v_{+} +\frac12 v_{+} \partial_z g + \frac12 v_{+} \partial_z
u+\partial_z^2 v_{+})\\
-\frac{z}{3}  u +C_{1}&= \frac{\varepsilon}{4} (u v_{+} v_{-} 
- (v_{+} \partial_z v_{-} -  v_{-} \partial_z v_{+}))\\
-\frac{z}{3}  g+C_{2} &=\frac{\varepsilon}{4}  (- g v_{+} v_{-}
-   \partial_z (v_{+} v_{-}))\, ,
\label{viscuvpmself}\end{split}
\end{equation}
where 
\[
v_{\pm} = v \pm f, \;\;K_{\pm}=K_1 \pm K_2,\;\;
\varepsilon=\varepsilon_1\, .
\]
First, we note that equations \eqref{viscuvpmself} can be made
independent of ${\varepsilon}$ through the substitution ${\cal S}$:
\begin{equation} {\cal S} : v_{\pm} \to (\varepsilon)^{-1/3}  v_{\pm} ,   \;\;K
u \to (\varepsilon)^{-1/3}  u , \;\;K  g \to (\varepsilon)^{-1/3}  g,
\;\;  z  \to (\varepsilon)^{1/3} z\, .
\label{Stran}\end{equation}
It is instructive to leave the equations \eqref{viscuvpmself} in the current
form as the change of variables  we will perform to arrive at the
Hamiltonian formalism will lead anyway to
canonical coordinates that are invariant under the above transformation 
${\cal S}$.

The equations \eqref{viscuvpmself}  are explicitly invariant under $F_2$:
\[ 
F_2 : v_{\pm}  \leftrightarrow \pm v_{\pm}, \; \varepsilon \to -\varepsilon, \; g \to g, u \to u
\]
From the last two equations of \eqref{viscuvpmself} we derive
\begin{equation}
\begin{split}
\frac{v_{-}^{\prime}}{v_{-}}
&= \frac12 (u-g) -\frac{2}{\varepsilon v_{+}v_{-}} \big[
(C_2-\frac{z g}{3})+(C_1-\frac{z u}{3}) \big]
\\
\frac{v_{+}^{\prime}}{v_{+}} 
&= -\frac12 (u+g) +\frac{2}{\varepsilon v_{+}v_{-}}
\big[ (C_1-\frac{z u}{3}) -(C_2-\frac{z g}{3})\big] \, .
\label{v+-eqs}\end{split}
\end{equation}
The first order derivatives for $u,g$ are:
\begin{equation}
\begin{split}
(z g)_z &= -\frac{z}{4} (v_{-}^2+v_{+}^2)
+ \frac{6}{ \varepsilon \,v_{+}\,v_{-}} \big[
(C_2-\frac{z}{3} g)^2- (C_1-\frac{z}{3} u)^2 \big]
+\frac{3}{4 } (K_{+}\,v_{+}+\,K_{-}\,v_{-})
\\
(z u)_z &= \frac{z}{4 }(v_{-}^2-v_{+}^2)
+\frac{3}{4 } (K_{+}\,v_{+}-K_{-}\,v_{-}) \, .
\label{guzzeqs}\end{split}
\end{equation}
Introducing for convenience
\[ 
G = z g, \quad U = z u\, ,
\]
we can rewrite equations  \eqref{v+-eqs},\eqref{guzzeqs}  as
\begin{equation}\label{v+-UG}
\begin{split}
v_{-}^{\prime} &= \frac{v_{-}}{2z} (U-G) -\frac{2}{\varepsilon v_{+}} \big[
(C_2-\frac{G}{3})+(C_1-\frac{U}{3}) \big]\, ,
\\
v_{+}^{\prime}
&=  -\frac{v_{+}}{2z} (U+G) -\frac{2}{\varepsilon v_{-}}
\big[ (C_2-\frac{G}{3})- (C_1-\frac{U}{3}) \big]\, ,\\
(G)_z &= -\frac{z}{4} (v_{-}^2+v_{+}^2)
+ \frac{6}{ \varepsilon \,v_{+}\,v_{-}} \big[
(C_2-\frac{1}{3} G)^2- (C_1-\frac{1}{3} U)^2 \big]
+\frac{3}{4 } (K_{+}\,v_{+}+\,K_{-}\,v_{-})\, ,
\\
(U)_z &= \frac{z}{4 }(v_{-}^2-v_{+}^2)
+\frac{3}{4 } (K_{+}\,v_{+}-K_{-}\,v_{-}) \, .
\end{split}
\end{equation}
There is one further change of variables needed to end up with equations that are
manifestly Hamilton equations, namely:
\[{\bar G}= \frac13 G -C_2, \quad {\bar U}= \frac13 U -C_1 \, .
\]
Equations for ${\bar G}, {\bar U}$ variables are:
\begin{equation}
\begin{split}
\left({\bar G}+{\bar U}\right)_z &= -\frac{z}{2 \cdot 3} v_{+}^2
+ \frac{2}{ \varepsilon \,v_{+}\,v_{-}} \left({\bar G}+{\bar U}\right)
\left({\bar G}-{\bar U}\right)
+\frac{1}{2 } K_{+}\,v_{+} \, ,
\\
\left({\bar G}-{\bar U}\right)_z 
&= -\frac{z}{2 \cdot 3} v_{-}^2
+ \frac{2}{ \varepsilon \,v_{+}\,v_{-}} \left({\bar G}+{\bar U}\right)
\left({\bar G}-{\bar U}\right)
+\frac{1}{2 } K_{-}\,v_{-} \, .
\label{barUG}\end{split}
\end{equation}
To end up with the polynomial Hamilton equations we further introduce :
\[
F_{+}=\frac{{\bar G}+ {\bar U}}{v_{+}}, \quad
F_{-}=\frac{{\bar G}- {\bar U}}{v_{-}}.
\]
Using this notation the first two of equations \eqref{v+-UG}
can be rewritten as:
\begin{equation}
v_{\pm}^{\prime} = -\frac{3}{2z} v_{\pm}^2 F_{\pm}
- \frac{3}{2z} v_{\pm} (C_2 \pm C_1)
+\frac{2}{\varepsilon } F_{\mp} \,.
\label{v+-F}
\end{equation}
From equations \eqref{barUG} and \eqref{v+-F} we obtain
\begin{equation}
\left(F_{\pm}\right)^{\prime} = - \frac{z}{3 \cdot 2} v_{\pm} +\frac12
K_{\pm} +\frac{3}{2z} v_{\pm} F_{\pm}^2 +\frac{3}{2z} (C_2\pm C_1)
F_{\pm}\, .
\label{1v+-}
\end{equation}
Define now the Hamiltonian :
\begin{equation}
\begin{split}
H&=\left(\frac{3}{4z} v_{+}^2 F_{+}^2  +\frac12
K_{+} v_{+} - \frac{z}{3 \cdot 4} v_{+}^2+ \frac{3}{2z} (C_1+C_2)
F_{+} v_{+} \right)- \frac{2}{\varepsilon} F_{+} F_{-}  \\
&+ \left(\frac{3}{4z} v_{-}^2 F_{-}^2  +\frac12
K_{-} v_{-} - \frac{z}{3 \cdot 4} v_{-}^2+ \frac{3}{2z} (C_1-C_2)
F_{-} v_{-} \right) \, .
\label{HamHam}
\end{split}
\end{equation}
which is polynomial in all variables such that it reproduces equations \eqref{v+-F}-\eqref{1v+-} through
\begin{equation} 
\left(F_{\pm}\right)^{\prime} = \frac{\delta}{\delta v_{\pm}} H,\quad
\left(v_{\pm}\right)^{\prime} = -\frac{\delta}{\delta F_{\pm}} H\, .
\label{Hameqsplusminus}
\end{equation}
Note that the ``plus'' and ``minus'' parts of H in \eqref{HamHam} are
connected by only one term  $- \frac{2}{\varepsilon} F_{+} F_{-}$.

The transformation
\[ F_{+} \to F_{+} + \frac{a}{v_{+}}, \;\; v_{+} \to v_{+}
\]
or
\[ v_{+} \to v_{+} + \frac{a}{F_{+}}, \;\; F_{+} \to F_{+}
\]
leaves only the first part of Hamiltonian \eqref{HamHam} invariant 
(up to a constant). 

We will now attempt to cast 
equations \eqref{v+-F}-\eqref{1v+-} in a form of equations that are
manifestly $D_4^{(1)}$ invariant \cite{sasanoD41}.

First, we apply the redefinition
\[ F_{\pm} \to F_{\pm} +\frac{z}{3},   \quad v_{\pm} \to  v_{\pm} \, ,
\]
to obtain
\begin{equation}
\begin{split}
\left(F_{+}\right)^{\prime} &= \frac{3}{2z} v_{+} F_{+}^2 + v_{+} F_{+}
  +\frac{3}{2z} (C_1+C_2) F_{+}+\left( \frac12 K_{+} +\frac{1}{2}
  (C_1+C_2)-\frac13\right)\, , \\
v_{+}^{\prime} &=  
-\frac{3}{2z} v_{+}^2 F_{+} - \frac12 v_{+}^2-\frac{3}{2z} v_{+} (C_1+C_2)
+\frac{2}{\varepsilon} (F_{-}+\frac{z}{3}) \, , 
\label{Fvplus}\end{split}
\end{equation}
and
\begin{equation}
\begin{split}
\left(F_{-}\right)^{\prime} &= \frac{3}{2z} v_{-} F_{-}^2
+v_{-} F_{-} -\frac{3}{2z} (C_1-C_2) F_{-}
 +\left( \frac12 K_{-} - \frac12 (C_1-C_2)-\frac13\right)\, ,  \\
v_{-}^{\prime} &= -\frac{3}{2z} v_{-}^2 F_{-}- \frac12 v_{-}^2
- \frac{3}{2z} v_{-} (C_2-C_1)
+\frac{2}{\varepsilon} (F_{+}+\frac{z}{3}) \, .
\label{Fvminus}\end{split}
\end{equation}
We now further substitute
\begin{equation}
F_{+} \to z F_{p}, \; F_{-} \to z F_{m}, \; v_{+} \to \frac{1}{z} v_{p}, \;
v_{-} \to \frac{1}{z} v_{m},
\label{subsFv}
\end{equation}
to obtain for $F_{p},v_{p}$ equations :
\begin{equation}
\begin{split}
\left(F_{p}\right)^{\prime} &= \frac{1}{z} \Big[ \frac{3}{2} v_{p} F_{p}^2 
+ v_{p} F_{p}
  +\frac{3}{2} (C_1+C_2-\frac23 ) F_{p}+\left( \frac12 K_{+} +\frac{1}{2}
  (C_1+C_2)-\frac13\right)\Big]\\
v_{p}^{\prime} &=\frac{1}{z} \Big[  
-\frac{3}{2} v_{p}^2 F_{p} - \frac12 v_{p}^2-\frac{3}{2} v_{p}
(C_1+C_2-\frac23 )\Big]
+\frac{2 z^2}{\varepsilon} (F_{m}+\frac{1}{3}) \, .
\label{Fvpp}\end{split}
\end{equation}
Introducing $\alpha_1 +\alpha_2 = (C_1+C_2-\frac23 )$
and $\alpha_2=\frac12 K_{+} +\frac{1}{2}
  (C_1+C_2)-\frac13$ we can rewrite the above equations as
\begin{equation}
\begin{split}
\left(F_{p}\right)^{\prime} &= \frac{1}{z} \Big[ \frac{3}{2} v_{p} F_{p}^2 
+ v_{p} F_{p}
  +\frac{3}{2} (\alpha_1+\alpha_2) F_{p}+\alpha_2
  \Big]\\
v_{p}^{\prime} &=\frac{1}{z} \Big[  
-\frac{3}{2} v_{p}^2 F_{p} - \frac12 v_{p}^2-\frac{3}{2}
(\alpha_1+\alpha_2) v_{p}
\Big] +\frac{2 z^2}{\varepsilon} (F_{m}+\frac{1}{3}) \, .
\label{Fvppp}\end{split}
\end{equation}  
For the ``$-$'' sector we obtain
\begin{equation}
\begin{split}
\left(F_{m}\right)^{\prime} &= \frac{1}{z} \Big[ \frac{3}{2} v_{m} F_{m}^2
+v_{m} F_{m} -\frac{3}{2} (C_1-C_2+\frac23 ) F_{m}
 +\left( \frac12 K_{-} - \frac12 (C_1-C_2)-\frac13\right)\Big] \\
v_{m}^{\prime} &= \frac{1}{z} \Big[-\frac{3}{2} v_{m}^2 F_{-}- \frac12
v_{m}^2
- \frac{3}{2} v_{m} (C_2-C_1 -\frac23 )\Big] 
+\frac{2 z^2 }{\varepsilon} (F_{p}+\frac{1}{3})\, .
\label{Fvmm}\end{split}
\end{equation}
Introducing $\alpha_3+\alpha_4 =- (C_1-C_2+\frac23 )$
and $\alpha_4=\frac12 K_{-} -\frac{1}{2}
 (C_1-C_2)-\frac13$ we can compactly rewrite the above equations as
\begin{equation}
\begin{split}
\left(F_{m}\right)^{\prime} &= \frac{1}{z} \Big[ \frac{3}{2} v_{m} F_{m}^2
+v_{m} F_{m} +\frac{3}{2} (\alpha_3+\alpha_4 ) F_{m}
 +\alpha_4 \Big] \\
v_{m}^{\prime} &= \frac{1}{z} \Big[-\frac{3}{2} v_{m}^2 F_{m}- \frac12
v_{m}^2
- \frac{3}{2} (\alpha_3+\alpha_4 ) v_{m} \Big] 
+\frac{2 z^2 }{\varepsilon} (F_{p}+\frac{1}{3}) \, .
\label{Fvmmm}\end{split}
\end{equation}
Equations \eqref{Fvppp}  and \eqref{Fvmmm} can be obtained from
the Hamiltonian:
\begin{equation}
\begin{split}
H&=\frac{1}{z} \left(\frac{3}{4} v_{p}^2 F_{p}^2  +\frac12
v_{p}^2 F_{p} + \frac{3}{2} (\alpha_1+\alpha_2)
F_{p} v_{p}  +\alpha_2 v_p  \right) -\frac{2 z^2}{\varepsilon} (F_m+\frac13) (F_p+\frac13)\\
&+\frac{1}{z} \left(\frac{3}{4} v_{m}^2 F_{m}^2  +\frac12
v_{m}^2 F_{m} + \frac{3}{2} (\alpha_3+\alpha_4)
F_{m} v_{m}  +\alpha_4 v_m  \right) \, .
\label{HamHams}
\end{split}
\end{equation}
through 
\begin{equation} 
\left(F_{i}\right)^{\prime} = \frac{\delta}{\delta v_{i}} H,\quad
\left(v_{i}\right)^{\prime} = -\frac{\delta}{\delta F_{i}} H, \;\; \;
i=p,m \, .
\label{Hameqspm}
\end{equation}
The author of \cite{sasanoD41} 
has proposed such system as two coupled Painlev\'e III equations
involving four variables and derived by symmetry consideration as a 
system that admits affine Weyl group symmetry of type $D_4^{(1)}$.

Comparing equations \eqref{Fvppp} and \eqref{Fvmmm} we notice presence
of $\pi_0$ automorphism : 
\[ \pi_0: \quad v_{p}, F_{p} \leftrightarrow v_{m}, F_{m}, \quad 
\alpha_2 \leftrightarrow \alpha_4,\; \alpha_1 \leftrightarrow 
\alpha_3 \, ,\]
that transforms equation \eqref{Fvppp} into \eqref{Fvmmm} and vice-versa.

In addition we introduce a variable  $\alpha_0$ 
defined by the 
condition $2 \alpha_0+\alpha_1+ \alpha_2+\alpha_3+\alpha_4={\rm const}$
\cite{sasanoD41}. 
The constant used to define $\alpha_0$ will be fixed below by a symmetry 
transformation $s_0$,  that mixes the ``$+/-$'' sectors to be defined below. 
In \cite{sasanoD41} 
that constant is set to $1$ consistently with Sasano's normalization (different from ours).

Furthermore we also find the following B\"acklund transformation
$s_2$ : 
\begin{equation}
s_2: \; v_p \to v_p + \frac{2 \alpha_2}{F_p}, \; F_p \to F_p , \;
\alpha_2 \to - \alpha_2\, ,
\label{s1back}
\end{equation}
that keeps equations \eqref{Fvppp} invariant. The consequence of 
$s_2(\alpha_2)=- \alpha_2$ is that $s_2(\alpha_0)= \alpha_0+ \alpha_2$
just to keep the 
condition $2 \alpha_0+\alpha_1+ \alpha_2+\alpha_3+\alpha_4={\rm const}$
unchanged.

Similarly the following B\"acklund transformation : 
\begin{equation}
s_4 :\; v_m \to v_m + \frac{2 \alpha_4}{F_m}, \; F_m \to F_m, \;
\alpha_4 \to - \alpha_4\, ,
\label{s4back}
\end{equation}
will keep equations \eqref{Fvmmm} invariant.

Note that $s_2^2=s_4^2= 1, \quad s_2 s_4= s_4 s_2$
and $\pi_0 s_2 \pi_0= s_4$. 

Furthermore, inspired by the automorphism \eqref{f2def}, we define the
two automorphisms:
\begin{equation}
\begin{split}
\pi_1 &: \; v_p \to - v_p, \; F_p \to -\frac{2}{3}-F_p , \; \varepsilon 
\to - \varepsilon,
\; \alpha_1 \to \alpha_2 \to \alpha_1 \\
\pi_3 &: \; v_m \to - v_m, \; F_m \to -\frac{2}{3}-F_m , \; \varepsilon  \to - 
\varepsilon ,
\; \alpha_3 \to \alpha_4 \to \alpha_3\, ,
\label{pi12}
\end{split}
\end{equation}
that both keep equations \eqref{Fvppp} - \eqref{Fvmmm} invariant
and satisfy
\[
\pi_1^2=\pi_3^2= 1, \quad \pi_0 \pi_1  \pi_0= \pi_3, \quad \pi_0  \pi_3 \pi_0 =\pi_1\, .
\]
Coincidently, all the canonical coordinates $v_p, v_m, F_p, F_m$ 
have been defined in such a way that they
are invariant under transformation ${\cal S}$ defined in 
relation \eqref{Stran}, while the substitution  $z  \to (\varepsilon)^{1/3} z$
allows to eliminate $\varepsilon$ completely 
from equations \eqref{Fvppp} - \eqref{Fvmmm}.
With $\varepsilon$ being replaced by $1$, one can alternatively define
the automorphisms $\pi_1, \pi_3$ involving a change of the sign of $z \to - z$ 
instead of $\varepsilon  \to - \varepsilon$, as it was done
in \cite{sasanoD41}.

The other two B\"acklund transformations are ( $s_0,s_3$ in notation of
\cite{sasanoD41}) but here relabeled as :
\begin{equation}
\begin{split}
s_1 &: \; v_p \to  v_p+\frac{2 \alpha_1}{F_p+\frac23 }, \; F_p \to F_p , \; 
\; \alpha_1 \to - \alpha_1, \; \alpha_2 \to \alpha_2 , \; \alpha_0 \to
\alpha_0+\alpha_1\\
s_3 &: \; v_m \to  v_m+\frac{2 \alpha_3}{F_m+\frac23 }, \; F_m \to F_m , \; 
\; \alpha_3 \to - \alpha_3, \; \alpha_4 \to \alpha_4 , \; \alpha_0 \to
\alpha_0+\alpha_3\, .
\label{s03}
\end{split}
\end{equation}
They both square to one : $ s_1^2=s_3^2= 1$. Also the B\"acklund
transformations satisfy :
\[
\begin{split}
\pi_i s_i \pi_i&=s_{i+1}, \;\;\;\;  \pi_i s_{i+1} \pi_i=s_{i},\;\;
\;\; i=1,3,\\
\pi_i s_{i \pm 2} \pi_i&=s_{i \pm 2} , \;\;\;\; 
\pi_i s_{i \pm 3} \pi_i=s_{i \pm 3} ,  \;\;\;\; +/- \;\;\;\;
{\rm for}\;\;\;\; i=1/3 \\
\pi_0 s_i \pi_0 &=s_{i + 2} , \;\;\;\; i=1,2
\end{split}
\]

Finally we need to prove invariance under $s_0$ that mixes the $+/-$
sectors. When this B\"acklund transformation is defined as
\begin{equation}
\begin{split}
s_0 (F_p) &= F_p - \frac{2 \alpha_0 \, v_m}{v_p v_m -\frac{4}{3
\varepsilon} z^3},
\;\; s_0 (v_p)=v_p, \; s_0 (\alpha_1)= \alpha_1+\alpha_0, \;
s_0 (\alpha_2)= \alpha_2+\alpha_0 \\
s_0(F_m) &= F_m - \frac{2 \alpha_0 \, v_p}{v_p v_m -\frac{4}{3 \varepsilon} z^3},
\;\; s_0 (v_m)=v_m, \; s_0 (\alpha_3)= \alpha_3+\alpha_0, \;
s_0 (\alpha_4)= \alpha_4+\alpha_0\\
s_0 (\alpha_0)&=-\alpha_0\, ,
\label{s02}
\end{split}
\end{equation}
the equations \eqref{Fvppp} and \eqref{Fvmmm} are invariant if the
condition,
\[
2 \alpha_0+\alpha_1+\alpha_2+\alpha_3+\alpha_4=-2 \, ,
\]
holds.
As remarked before our normalization is different from the one used by Sasano
\cite{sasanoD41} and the differences also include different powers of
$z$ in equations \eqref{s02} and in the Hamiltonian \eqref{HamHams}.

Note that $s_0^2=1$ because ${\bar \alpha_0}=-\alpha_0$ and $\pi_0 s_0
\pi_0=s_0$, $\pi_i s_0 \pi_i,\, i=1,3$.
It is easy to verify that the  $D_4^{(1)}$ $s_i, \,1,2,3,4$ 
B\"acklund transformations satisfy :
\[
\begin{split}
s_i^2&=1 , \quad i=1,2,3,4\, ,\\
s_i s_j&=s_j s_i, \quad i,j=1,2,3,4\, ,\\
s_i s_0 s_i &=s_0 s_i s_0, \quad i=1,2,3,4\, ,
\end{split}
\]
where the last two identities are equivalent to the standard 
$D_4^{(1)}$ relations $(s_i s_j)^2=1, (s_0 s_i)^3=1$. This is in
contrast to the $A^{(1)}_l$ affine Weyl symmetry group multiplications
for which it holds that  $(s_i s_{i \pm 1})^3=1$ and $(s_i s_{i \pm
2})^2=1$. These examples clearly illustrate
a difference from the $D_4^{(1)}$ structure encountered above.

The steps shown in this section complete the systematic derivation of the $D_4^{(1)}$
Hamiltonian system starting from the integrable hierarchy of $D_4$
symmetry. We will return to the model with two independent parameters
$\varepsilon_i, i=1,2$ in a separate publication.

This work illustrates the power of algebraic methods to
derive systems invariant under affine Weyl groups that should lend itself
well to generalizations to other group structures.

Recently, the Sasano systems of four-dimensional Painlev\'e 
type equations  with affine Weyl group symmetry of 
type  $D_6^{(1)}$ \cite{sasanorims} were derived 
as isomonodromic deformation equations in \cite{sakai,kawakami}, which
suggests that a similar analysis will apply to coupled Painlev\'e III models 
with four canonical variables of reference \cite{sasanoD41} obtained in 
this paper from the self-similarity
limit.

\appendix
\section{Algebraic background on $so(2n)$ } 
\label{section:algebraic}
Here we discuss the Lie algebra so(2n) and its loop algebra that
underlies the zero-curvature considerations.
The algebra $so(2n)=\{ X \in {gl}(2n,C) \vert X+X^T=0\}$ is generated by $2n \times 2n$ anti-symmetric matrices 
$ L_{i, j}=-L_{j ,i}$  with components
\begin{equation}
(L_{i,j})_{k,l} = \delta_{il} \delta_{jk}-\delta_{ik} \delta_{jl}, \quad
i,j=1,{\ldots} ,2n\, .
\label{defL}
\end{equation}
These $\frac12 (2 n )(2 n - 1)$ matrices form a basis for the $so(2n)$
Lie algebra with the  commutation relations :
\begin{equation}
\lbrack L_{i,j}, L_{m,n} \rbrack = \delta_{i,m} L_{j,n}+\delta_{j,n} L_{i,m}
-\delta_{i,n} L_{j,m} -\delta_{j,m}L_{i,m} \, ,
\label{commutator}
\end{equation}
where we followed Olive's convention \cite{olive}. The Cartan sub-algebra 
generators are:
\[
H_i= i L_{2i-1,2 i} ,\quad 1,2,{\ldots}, n \, .
\]
The relevant commutation relations in accordance to \eqref{commutator}
are:
\begin{equation}
\begin{split}
\lbrack H_{i}, L_{2j-1, 2k-1} \rbrack &=  i \delta_{i,j} L_{2j, 2k-1}-i
\delta_{i,k} L_{2i, 2j-1} \\
\lbrack H_{i}, L_{2j, 2k-1} \rbrack &=  -i \delta_{i,j} L_{2j-1, 2k-1}+i
\delta_{i,k} L_{2j,2k} \\
\lbrack H_{i}, L_{2j-1, 2k} \rbrack &=  i \delta_{i,j} L_{2j, 2k}-i
\delta_{i,k} L_{2j-1,2k-1} \\
\lbrack H_{i}, L_{2j, 2k} \rbrack &=  -i \delta_{i,j} L_{2j-1, 2k}-i
\delta_{i,k} L_{2j,2k-1} 
\label{HL2j2k}
\end{split}
\end{equation}
The roots are:
\[
 {\boldsymbol \alpha}= \epsilon \,  {\mathbf e}_j +\eta \, {\mathbf e}_k, \quad j \ne k
\]
with independent $\epsilon, \eta = \pm 1$ and 
$e_i, i=1,{\ldots} ,n$ with $(e_i,e_j)=\delta_{i,j}$ being a basis for
$R^n$.

The associated step operators are
\[
E_{\boldsymbol \alpha} =-\frac12 \left( L_{2j-1,2k-1}+ i \epsilon   L_{2j,2k-1}
+i \eta  L_{2j-1,2k} - \epsilon \eta L_{2j,2k} \right)
\]
Number of roots is $\frac12 n (n-1) \times 2 \times2= 2 n(n-1)$,which is
the dimension of $so(2n)$, rank of Cartan sub-algebra is $n$.

It holds that
\[
\lbrack H_{i}, E_{\boldsymbol \alpha}  \rbrack = (\epsilon
\delta_{i,j} +
\eta \delta_{i,k} ) E_{\boldsymbol \alpha} \, ,
\]
as long as $\eta^2=1,\epsilon^2=1$.

All roots have equal length and satisfy
$ ( {\boldsymbol \alpha}, {\boldsymbol \alpha})=2$.
The basis of simple roots is given by:
\begin{equation}
\begin{split}
{\boldsymbol \alpha}_i &=  {\mathbf e}_i-  {\mathbf e}_{i+1}, \;
i=1,{\ldots} ,n-1\\
{\boldsymbol \alpha}_n &=  {\mathbf e}_{n-1}+ {\mathbf e}_{n}, 
\label{simpleroots}
\end{split}
\end{equation}
The inner product of simple roots
\[
( {\boldsymbol \alpha}_i, {\boldsymbol \alpha}_j)= 
\left\{
\begin{matrix}
2 & i=j & 1 \le i,j \le n\\
-1 & \vert i-j \vert =1& 1 \le i,j \le n-1\\
0&\vert i-j \vert \ge 2& 1 \le i,j \le n\\
0& i=n-1, j=n&
\end{matrix}
\right.\
\]
defines the corresponding Cartan matrix.
For $so(2n)$ the roots and co-roots are identical, the highest root is
\[ \psi= e_1 +e_2= {\boldsymbol \alpha}_1+2 {\boldsymbol \alpha}_2+{\ldots} +2 {\boldsymbol \alpha}_{n-2}
+{\boldsymbol \alpha}_{n-1} +{\boldsymbol \alpha}_{n}\, , 
\]
the Coxeter number $h$  and the dual Coxeter number $h^\vee$ coincide and
\[h=h^\vee=2n-2 \,. \]
For case of $so(8)$ these become 
\begin{equation} \psi= e_1 +e_2={\boldsymbol \alpha}_1+2 {\boldsymbol \alpha}_2+
+{\boldsymbol \alpha}_{3} +{\boldsymbol \alpha}_{4}, \qquad
h=h^\vee=6
\label{highroot8}
\end{equation}
The fundamental weights $\Lambda_i$ such that $2 ({\boldsymbol \alpha}_i,\Lambda_j)/
({\boldsymbol \alpha}_i,{\boldsymbol \alpha}_i)=\delta_{ij}$ are:
\[ \Lambda_i= \sum_{j=1}^i e_j = {\boldsymbol \alpha}_1+2 {\boldsymbol \alpha}_2 + {\ldots}  +
(i-1) {\boldsymbol \alpha}_{i-1} + i ( {\boldsymbol \alpha}_i+ {\ldots} + {\boldsymbol \alpha}_{n-2})+
\frac{i}{2}({\boldsymbol \alpha}_{n-1}+{\boldsymbol \alpha}_n), \; i=1,{\ldots} ,n-2\]
\[
\Lambda_{n-1}= \frac12 (e_1+{\ldots} +e_{n-1}-e_n) = \frac12 ({\boldsymbol \alpha}_1+
2 {\boldsymbol \alpha}_2 + {\ldots}  + (n-2) {\boldsymbol \alpha}_{n-2}) +\frac{n}{2} {\boldsymbol \alpha}_{n-1}
+ \frac{n-2}{2}{\boldsymbol \alpha}_n
\]
\[
\Lambda_{n}= \frac12 (e_1+{\ldots} +e_{n-1}+e_n) = \frac12 ({\boldsymbol \alpha}_1+
2 {\boldsymbol \alpha}_2 + {\ldots}  + (n-2) {\boldsymbol \alpha}_{n-2}) +\frac{n-2}{2} {\boldsymbol \alpha}_{n-1}
+ \frac{n}{2}{\boldsymbol \alpha}_n
\]
Especially for   $so(8)$ with $n=4$ we find for weights and simple roots
\[ 
\begin{split}
\Lambda_1&= e_1 , \qquad\qquad \qquad\qquad \quad\;\;
{\boldsymbol \alpha}_1 = e_1-e_2 ,\\
\Lambda_2 &= e_1+e_2 , \qquad\qquad\qquad\quad\;\; 
{\boldsymbol \alpha}_2 = e_2-e_3, \\
\Lambda_{3}& = \frac12 (e_1+e_2 +e_3-e_4) , \qquad {\boldsymbol
\alpha}_3 = e_3-e_4,\\
\Lambda_{4}& = \frac12 (e_1+e_2 +e_3+e_4) , \qquad {\boldsymbol
\alpha}_4 = e_3+e_4,
\end{split}
\]
we obtain for a sum of weights:
\begin{equation}
\Lambda= \sum_{i=1}^4 \Lambda_i= 3 e_1 +2 e_2 +e_3
\label{Lambda}
\end{equation}
The product of $\Lambda$ and a general root ${\boldsymbol\alpha}
= \epsilon e_i+\eta e_j$
\[
( \Lambda, {\boldsymbol\alpha}) \ne 0
\]
for all ${\boldsymbol\alpha}= \epsilon e_i+\eta e_j$.

We will use \eqref{Lambda} to define the principal gradation operator 
for $so(8)$:
\begin{equation}
Q= 6 d + \sum_{i=1}^n \Lambda_i\cdot H
=6 d +\Lambda\cdot H = 6 d +\left( 3 e_1 +2 e_2 +e_3\right) \cdot H \, .
\label{qprincipal}
\end{equation}
Note that 
\[
(\Lambda, \psi)= \left( 3 e_1 +2 e_2 +e_3, e_1 +e_2\right)=5 \, .
\]

\subsection{so(8) charge sectors and their bases }
\label{subsection:sectors}

The underlying charge sectors  are (with $m \in  \mathbb{Z}$):
\begin{equation}\begin{split}
{\cal G}^{(6 m)}&= \{ H_1^{(m)}, H_2^{(m)}, H_3^{(m)}, H_4^{(m)} \}\\
{\cal G}^{(6m+1)}&= \{ E^{(m)}_{e_1-e_2}, E^{(m)}_{e_2-e_3},
E^{(m)}_{e_3-e_4}, E^{(m)}_{e_3+e_4} ,  E^{(m+1)}_{-\psi}=
E^{(m+1)}_{-e_1-e_2}\}\\
{\cal G}^{(6m+2)}&= \{ E^{(m)}_{e_1-e_3},  E^{(m)}_{e_2+e_4},
E^{(m)}_{e_2-e_4},  E^{(m+1)}_{-e_1-e_3} \}\\
{\cal G}^{(6m+3)}&=\{E^{(m)}_{e_1-e_4}, E^{(m)}_{e_1+e_4},
E^{(m)}_{e_2+e_3},E^{(m+1)}_{-e_1-e_4},  E^{(m+1)}_{-e_1+e_4},
E^{(m+1)}_{-e_2-e_3} \}\\
{\cal G}^{(6m+4)}&=\{ E^{(m)}_{e_1+e_3},
E^{(m+1)}_{-e_2+e_4},E^{(m+1)}_{-e_2-e_4},  E^{(m+1)}_{-e_1+e_3}\}\\
{\cal G}^{(6m+5)}&= \{E^{(m)}_{e_1+e_2}, E^{(m+1)}_{-e_3+e_4} ,
E^{(m+1)}_{-e_3-e_4},  E^{(m+1)}_{-e_2+e_3},  E^{(m+1)}_{-e_1+e_2} \}\, .
\label{chargesectors}
\end{split}\end{equation}

The unique grade one semi-simple element in ${\cal G}^{(1)}$ is
\begin{equation}
\begin{split}
E^{(1)}&= \sum_{i=1}^4 E^{(0)}_{\alpha_i}+ E^{(1)}_{-\psi}\\
&= E^{(0)}_{e_1-e_2}+ E^{(0)}_{e_2-e_3}+
E^{(0)}_{e_3-e_4}+ E^{(0)}_{e_3+e_4}+ E^{(1)}_{-e_1-e_2}
\label{egradeone}
\end{split}\end{equation}
where the sum was over all simple roots of $so(8)$ from \eqref{simpleroots}.

Define the kernels ${\cal K}^{(i)} \in {\cal G}^{(i)}$
to be such that
\[ \lbrack E^{(1)}, {\cal K}^{(i)} \rbrack=0\, ,
\]
for $i=2,3,4,5$ and ${\cal G}^{(i)}$ as given in \eqref{chargesectors}.

Given the grade $2$ sector ${\cal G}^{(2)}$ in \eqref{chargesectors}
we consider 
\begin{equation}
\begin{split}
\lbrack E^{(1)}, E^{(0)}_{e_1-e_3} \rbrack&= - E^{(0)}_{e_1-e_4}-
 E^{(0)}_{e_1+e_4}- E^{(1)}_{-e_2-e_3} \\
\lbrack E^{(1)}, E^{(0)}_{e_2+e_4} \rbrack &= + E^{(0)}_{e_1+e_4}+
 E^{(0)}_{e_2+e_3}+ E^{(1)}_{-e_1+e_4}\\
\lbrack E^{(1)}, E^{(0)}_{e_2-e_4} \rbrack &= + E^{(0)}_{e_1-e_4}+
 E^{(0)}_{e_2+e_3}+ E^{(1)}_{-e_1-e_4}\\
 \lbrack E^{(1)}, E^{(1)}_{-e_1-e_3} \rbrack &= - E^{(1)}_{-e_2-e_3}-
 E^{(1)}_{-e_1-e_4}- E^{(1)}_{-e_1+e_4}
\label{EEpe1me3}
\end{split}
\end{equation}

Accordingly we find for 
\[
K^{(2)}= a E^{(0)}_{e_1-e_3} + b  E^{(0)}_{e_2+e_4} +c E^{(0)}_{e_2-e_4}
+d  E^{(1)}_{-e_1-e_3} 
\]
that 
\[ \lbrack E^{(1)}, K^{(2)} \rbrack = 0
\]
only for $a=b=c=d=0$  and ${\cal K}^{(2)}$ is empty.

For 
\[
K^{(3)}= \varepsilon_1 E^{(0)}_{e_1-e_4} + \varepsilon_2  E^{(0)}_{e_1+e_4} +\varepsilon_3
E^{(0)}_{e_2+e_3}
+\varepsilon_4  E^{(1)}_{-e_1-e_4}+ \varepsilon_5  E^{(1)}_{-e_1+e_4}+\varepsilon_6 E^{(1)}_{-e_2-e_3} 
\]
we find that
\begin{equation}
\lbrack E^{(1)}, K^{(3)} \rbrack = 0 \quad \text{for} \quad \varepsilon_3=-\varepsilon_1-\varepsilon_2,
\varepsilon_6=-\varepsilon_1-\varepsilon_2,\varepsilon_4=\varepsilon_2,\varepsilon_5=\varepsilon_1
\label{e1k3}
\end{equation}
with arbitrary two parameters $\varepsilon_1,\varepsilon_2$ that parameterize ${\cal K}^{(3)}$.
If we define elements in  ${\cal K}^{(3)}$ that both satisfy \eqref{e1k3}:
\begin{equation}\begin{split}
K^{(3)}_\varepsilon&= \varepsilon_1 E^{(0)}_{e_1-e_4} + \varepsilon_2  E^{(0)}_{e_1+e_4} -(\varepsilon_1+\varepsilon_2)
E^{(0)}_{e_2+e_3}
+\varepsilon_2  E^{(1)}_{-e_1-e_4}+ \varepsilon_1  E^{(1)}_{-e_1+e_4}-(\varepsilon_1+\varepsilon_2)
E^{(1)}_{-e_2-e_3} \\
K^{(3)}_\eta&= \eta_1 E^{(0)}_{e_1-e_4} + \eta_2  E^{(0)}_{e_1+e_4} -(\eta_1+\eta_2)
E^{(0)}_{e_2+e_3}
+\eta_2  E^{(1)}_{-e_1-e_4}+ \eta_1  E^{(1)}_{-e_1+e_4}-(\eta_1+\eta_2)
E^{(1)}_{-e_2-e_3} \, ,
\label{Kthree}
\end{split}
\end{equation}
then 
\[ \lbrack K^{(3)}_\varepsilon, K^{(3)}_\eta \rbrack=0\, ,
\]
for any two arbitrary sets $(\varepsilon_1,\varepsilon_2),(\eta_1,\eta_2)$. 
Thus ${\cal K}^{(3)}$ is, as expected, abelian.

For 
\[
K^{(4)}= \varepsilon_1 E^{(0)}_{e_1+e_3} + \varepsilon_2  E^{(1)}_{-e_2+e_4} +\varepsilon_3
E^{(1)}_{-e_2-e_4}
+\varepsilon_4  E^{(1)}_{-e_1+e_3} \, ,
\]
we find that
\begin{equation}
\lbrack E^{(1)}, K^{(4)} \rbrack = 0 \quad \text{for} \quad \varepsilon_1=0,
\varepsilon_2=0,\varepsilon_3=0,\varepsilon_4=0 \, ,
\label{e1k4}
\end{equation}
and ${\cal K}^{(4)}$ is empty. 

For 
\[
K^{(5)}= a E^{(0)}_{e_1+e_2} + b  E^{(1)}_{-e_3+e_4} +c E^{(1)}_{-e_3-e_4}
+d  E^{(1)}_{-e_2+e_3}+ e  E^{(1)}_{-e_1+e_2} \, ,
\]
we find that
\begin{equation}
\lbrack E^{(1)}, K^{(5)} \rbrack = 0 \quad \text{for} \quad b=a,
c=a,e=a,d=2a\, ,
\label{e1k5}
\end{equation}
with an arbitrary one parameter $a$ that parametrizes ${\cal K}^{(5)}$.

For the (sub-algebras) ${\cal G}^{(3)}$ and ${\cal G}^{(1)}$ that have
non-trivial two- and one-dimensional kernels,  ${\cal K}^{(3)}$ and
${\cal K}^{(1)}$, respectively,  it is useful to describe their bases.

For ${\cal G}^{(3)}$ from the relation \eqref{chargesectors} we will use the 
basis:
\begin{equation}\begin{split}
V_1&= E^{(0)}_{e_1-e_4} -E^{(0)}_{e_2+e_3}+ E^{(1)}_{-e_1+e_4}
-E^{(1)}_{-e_2-e_3}\, , \\
V_2&=  E^{(0)}_{e_1+e_4}-E^{(0)}_{e_2+e_3}+E^{(1)}_{-e_1-e_4}-E^{(1)}_{-e_2-e_3} 
\, ,\\
V_3&= E^{(0)}_{e_1-e_4}+E^{(0)}_{e_1+e_4}+E^{(1)}_{-e_2-e_3}\, , \\
V_4&= -E^{(0)}_{e_1+e_4} - E^{(0)}_{e_2+e_3}-E^{(1)}_{-e_1+e_4}\, ,\\
V_5&=-E^{(0)}_{e_1-e_4}-E^{(0)}_{e_2+e_3}- E^{(1)}_{-e_1-e_4}\, ,\\
V_6&=E^{(1)}_{-e_1-e_4} +E^{(1)}_{-e_1+e_4}+E^{(1)}_{-e_2-e_3} \, ,
\label{basisG3}
\end{split}
\end{equation}
with $V_1, V_2$ being the two matrices from
\eqref{e1k3} that span a basis for the kernel ${\cal K}^{(3)}$  of $E^{(1)}$ in ${\cal
G}^{(3)}$, while $V_3, V_4, V_5, V_6$ span a basis for  the image
of $E^{(1)}$ in ${\cal G}^{(3)}$.

To analyze zero-curvature equations involving ${\cal G}^{(1)}$ 
from the relation \eqref{chargesectors} we will use the basis 
$E_1, {\ldots} , E_5$ :
\begin{equation}\begin{split}
E_1 &= E^{(1)}, \;\; E_2= - E^{(0)}_{e_1-e_2}+E^{(1)}_{-e_1-e_2}\\
E_3&= E^{(0)}_{e_1-e_2}- E^{(0)}_{e_2-e_3}+ E^{(1)}_{-e_1-e_2}, \;\;
E_4= E^{(0)}_{e_2-e_3}-E^{(0)}_{e_3-e_4}-E^{(0)}_{e_3+e_4}\\
E_5&= E^{(0)}_{e_3-e_4}-E^{(0)}_{e_3+e_4}\,,
\label{g1basis}
\end{split}
\end{equation}
for ${\cal G}^{(1)}$ . The first
element $E_1$ is obviously in kernel of $ E^{(1)}$, while $E_2,E_3,E_4,E_5$
span the image of  $ E^{(1)}$. 
One can check that 
\[
\varepsilon_1 E_1 + \varepsilon_2 E_2 + \varepsilon_3 E_3 + \varepsilon_4 E_4 + \varepsilon_5 E_5=0 \;\; \to \;\;
\varepsilon_1=\varepsilon_2=\varepsilon_3=\varepsilon_4=\varepsilon_5=0\,
,
\]
and the same basic relation for the $V$-basis.

\section{Main expressions of of the zero-curvature calculation}
\label{section:curvaturanula}

The coefficients $M_i, i=1,{\ldots} ,4$ of the matrix  $D^{(2)}$
defined in  expressions \eqref{DtwoM},
are explicitly given by solving the grade $3$ equation
\eqref{grade3eq}:
\begin{equation}
\begin{split}
M_1 &= \frac{(\varepsilon_1+\varepsilon_2)}{2} ( \phi_1+ \phi_2+\phi_3)
-\frac{(\varepsilon_1-\varepsilon_2)}{2} \phi_4\, , \\
M_2 &= \frac{(\varepsilon_1+\varepsilon_2)}{2} (-\phi_4+\phi_2+\phi_3) +\frac{(\varepsilon_1-\varepsilon_2)}{2}
\phi_1\, , \\
M_3 &= \frac{(\varepsilon_1+\varepsilon_2)}{2} (\phi_2+\phi_3+\phi_4)-\frac{(\varepsilon_1-\varepsilon_2)}{2}
\phi_1\, , \\
M_4 &= \frac{(\varepsilon_1+\varepsilon_2)}{2} (-\phi_1+\phi_2+
\phi_3)+\frac{(\varepsilon_1-\varepsilon_2)}{2} \phi_4 \,.
\label{Mphis}
\end{split}
\end{equation}
The coefficients $d_i,i=2,{\ldots} ,5$ of the grade one 
element $D^{(1)}$ along the basis elements $E_i, i=2,{\ldots} E_5$ given in
expressions \eqref{g1basis}
are obtained from the grade $2$ component of the zero curvature equations 
\eqref{zerocurva} to be
\begin{equation}
\begin{split}
d_2&=-\frac{(\varepsilon_1+\varepsilon_2)}{2} \left( \phi_1\phi_2+\partial_x \phi_1\right)
-\frac{(\varepsilon_1-\varepsilon_2)}{2} \left( \phi_3\phi_4 -\partial_x \phi_4 \right)\, , \\
d_3 &= \frac{(\varepsilon_1+\varepsilon_2)}{6} \left(- \phi_2 \phi_3 +\phi_2^2 -2\phi_3^2-\phi_4^2
+2 \phi_1^2+ 3 \partial_x (\phi_2+\phi_3)\right)
-\frac{(\varepsilon_1-\varepsilon_2)}{6}  \phi_1\phi_4 \, , \\
d_4 &= \frac{(\varepsilon_1+\varepsilon_2)}{6} \left( \phi_2 \phi_3-\phi_3^2 +\phi_1^2-2 \phi_4^2
+2 \phi_2^2 +3 \partial_x (\phi_2+\phi_3)\right)
+\frac{(\varepsilon_1-\varepsilon_2)}{6} \phi_1\phi_4\, , \\
d_5 &= \frac{(\varepsilon_1+\varepsilon_2)}{2} \left( \phi_3\phi_4-\partial_x \phi_4 \right)
+\frac{(\varepsilon_1-\varepsilon_2)}{2} \left( \phi_1\phi_2+\partial_x \phi_1\right)
\label{dphis}
\end{split}
\end{equation}
The components of $\lbrack A_0 , D^{(1)} \rbrack=\sum_{i=2}^5 C_i E_i$
can be calculated as
\begin{equation} \begin{split}
C_2 &= -\phi_1  d_1-\phi_2 d_2-\phi_1d_3\, , \\
C_3 &= -\phi_2 d_1 -\frac23 \phi_1 d_2  -\frac13 \phi_2 d_3 -\frac13
\phi_3 d_3 -\frac13  \phi_2 d_4+\frac23 \phi_3 d_4+\frac13 \phi_4 d_5\, , \\
C_4 &=-\phi_3 d_1   - \frac13 \phi_1 d_2 - \frac23 \phi_2 d_3 +
\frac13 \phi_3 d_3 + \frac13 \phi_2 d_4 + \frac13 \phi_3 d_4 + \frac23
\phi_4 d_5\, , \\
C_5  &= -\phi_4 d_1+\phi_4 d_4+\phi_3 d_5 \, .
\label{newCia}
\end{split}
\end{equation}
Inserting these values of $d_i$ and $C_i$ into equation \eqref{vidici} 
we obtain 
\begin{equation}
\begin{split}
v_1 &= \frac{(\varepsilon_1+\varepsilon_2)}{4} (-\phi_1(\phi_2^2+\phi_3^2+\phi_4^2
-\phi_1^2)
+2 \phi_1\partial_x (\phi_3+2 \phi_2)+2\partial_x^2 \phi_1) \\
&+\frac{(\varepsilon_1-\varepsilon_2)}{2}( -\phi_2\phi_3\phi_4+\phi_2\partial_x \phi_4+\partial_x (\phi_3\phi_4)
-\partial_x^2 \phi_4)
\, ,\\
v_2 &= \frac{(\varepsilon_1+\varepsilon_2)}{4} (-\phi_2(\phi_1^2+\phi_3^2+\phi_4^2-\phi_2^2) 
+2 \phi_4\partial_x \phi_4 +2 (\phi_2 +\phi_3)\partial_x \phi_3
-4\phi_1\partial_x \phi_1
-2\partial_x^2 (\phi_2+ \phi_3))\\
&-\frac{(\varepsilon_1-\varepsilon_2)}{2}(\phi_1 \phi_3 \phi_4-\phi_1\partial_x
\phi_4)\, ,\\
v_3 &=\frac{(\varepsilon_1+\varepsilon_2)}{4} (-\phi_3(\phi_1^2+\phi_2^2+\phi_4^2-\phi_3^2)
+4 \phi_4\partial_x \phi_4-2(\phi_2+ \phi_3) \partial_x \phi_2
-2\phi_1\partial_x \phi_1-2\partial_x^2(\phi_2+ \phi_3) 
)\\
&-\frac{(\varepsilon_1-\varepsilon_2)}{2}( \phi_1\phi_2 \phi_4+\phi_4\partial_x \phi_1)
\, ,\\
v_4 &= \frac{(\varepsilon_1+\varepsilon_2)}{4} (-\phi_4(\phi_1^2+\phi_2^2+\phi_3^2-\phi_4^2)
-2 \phi_4\partial_x (\phi_2+2 \phi_3)+2\partial_x^2 \phi_4)\\
&-\frac{(\varepsilon_1-\varepsilon_2)}{2}(\phi_1\phi_2 \phi_3 +\phi_3\partial_x \phi_1+\partial_x (\phi_1\phi_2)
+\partial_x^2 \phi_1)\, .
\label{vis}
\end{split}
\end{equation}
We can now insert the above values $v_i$ into the $t_3$ flow
expression \eqref{t3flows}
to obtain 
\begin{equation}
\begin{split}
\partial_{t_3} \phi_1 &= \partial_x \big[ \frac{(\varepsilon_1+\varepsilon_2)}{4} (-\phi_1(\phi_2^2+\phi_3^2+\phi_4^2
-\phi_1^2)
+2 \phi_1\partial_x (\phi_3+2 \phi_2)+2\partial_x^2 \phi_1)\big]\\
&+\partial_x \big[ \frac{(\varepsilon_1-\varepsilon_2)}{2}( -\phi_2\phi_3\phi_4+\phi_2\partial_x \phi_4+\partial_x (\phi_3\phi_4)
-\partial_x^2 \phi_4)\big]
\, , \\
\partial_{t_3} \phi_2 &= \partial_x \big[ \frac{(\varepsilon_1+\varepsilon_2)}{4}
(-\phi_2(\phi_1^2+\phi_3^2+\phi_4^2-\phi_2^2) 
+2 \phi_4\partial_x \phi_4 +2 (\phi_2 +\phi_3)\partial_x \phi_3
-4\phi_1\partial_x \phi_1
-2\partial_x^2 (\phi_2+ \phi_3)\big] \\
&-\partial_x \big[ \frac{(\varepsilon_1-\varepsilon_2)}{2}( \phi_3\phi_1\phi_4-\phi_1\partial_x
\phi_4))\big] \, , \\
\partial_{t_3} \phi_3 &=\partial_x \big[ \frac{(\varepsilon_1+\varepsilon_2)}{4}
(-\phi_3(\phi_1^2+\phi_2^2+\phi_4^2-\phi_3^2)
+4 \phi_4\partial_x \phi_4-2(\phi_2+ \phi_3) \partial_x \phi_2
-2\phi_1\partial_x \phi_1-2\partial_x^2(\phi_2+ \phi_3) )
\big]\\
&-\partial_x \big[\frac{(\varepsilon_1-\varepsilon_2)}{2}(
\phi_2\phi_1\phi_4+\phi_4\partial_x \phi_1) \big]
\, , \\
\partial_{t_3} \phi_4 &= \partial_x \big[ \frac{(\varepsilon_1+\varepsilon_2)}{4} (-\phi_4
(\phi_2^2+\phi_3^2+\phi_1^2-\phi_4^2)
-2 \phi_4\partial_x (\phi_2+2 \phi_3)+2\partial_x^2 \phi_4)\big] \\
&+\partial_x \big[\frac{(\varepsilon_1-\varepsilon_2)}{2}(-\phi_1\phi_3\phi_2 -\phi_3\partial_x \phi_1-\partial_x (\phi_1\phi_2)
-\partial_x^2 \phi_1) \big] \, .
\label{visb}
\end{split}
\end{equation}

\subsection*{Acknowledgements}
This study was financed in part  by the Coordena\c{c}\~{a}o de
Aperfei\c{c}oamento de Pessoal de N\'ivel Superior - Brasil (CAPES) - Finance Code 001
(G.V.L.) and by CNPq and FAPESP (J.F.G. and A.H.Z.).
H.A. thanks Victor C. Alves for enouragement to embark on this study.

\label{lastpage}

\begin{thebibliography}{99}
\bibitem{sasanoD41}
Y. Sasano, Symmetries in the system of type $D_4^{(1)}$, arXiv:0704.2331
\bibitem{gerdjikov}
 V.S. Gerdjikov, A.A. Stefanov, I.D. Iliev  et al.,
Recursion operators and hierarchies of mKdV equations 
related to the Kac-Moody algebras 
$D_4^{(1)}$, $D_4^{(2)}$, $D_4^{(3)}$. 
{\it Theor Math Phys} {\bf 204}, 1110-1129 (2020). 
https://doi.org/10.1134/S0040577920090020
\bibitem{FS}
K. Fuji and T. Suzuki, 
The sixth Painlev\'e equation arises from a Drinfeld-Sokolov hierarchy of 
type $D_4^{(1)}$ by similarity reduction, {\it Journal of Physics A:
Mathematical and Theoretical} {\bf 39} 12073-12082 (2006).
\bibitem{olive}
D. I. Olive, Lectures on Gauge Theories and Lie Algebras, Univ. Virginia 
preprint 
\bibitem{sasanorims}
Y. Sasano,
{\it RIMS K\^oky\^uroku Bessatsu} {\bf B5} 137-152   (2008) 
\bibitem{sakai}
H. Sakai, 
{\it MSJ Memoirs} {\bf 37}  1-23 (2018)
\bibitem{kawakami}
H. Kawakami, {\it Journal of Integrable Systems} {\bf 3} 1-36 (2018) 

\end{thebibliography}
\end{document}